# Neutrino Physics at a Muon Collider

Bruce J. King

*Brookhaven National Laboratory*
*email: bking@bnl.gov*

**Abstract.** An overview is given of the neutrino physics potential of future muon storage rings that use muon collider technology to produce, accelerate and store large currents of muons.

## INTRODUCTION

This paper gives an overview of the neutrino physics possibilities at a future muon storage ring, which can be either a muon collider ring or a ring dedicated to neutrino physics that uses muon collider technology to store large muon currents.

After a general characterization of the neutrino beam and its interactions, some crude quantitative estimates are given for the physics performance of a muon ring neutrino experiment (MURINE) consisting of a high rate, high performance neutrino detector at a 250 GeV muon collider storage ring.

The paper is organized as follows. The next section describes neutrino production from a muon storage rings and gives expressions for event rates in general purpose and long baseline detectors. This is followed by a section outlining a serious design constraint for muon storage rings: the need to limit the radiation levels produced by the neutrino beam. The following two sections describe a general purpose detector and the experimental reconstruction of interactions in the neutrino target then, finally, the physics capabilities of a MURINE are surveyed.

## NEUTRINO PRODUCTION AND EVENT RATES

Neutrinos are emitted from the decay of muons in the collider ring:

$$\begin{aligned} \mu^- &\to \nu_\mu + \overline{\nu_e} + e^-, \\ \mu^+ &\to \overline{\nu_\mu} + \nu_e + e^+. \end{aligned} \quad (1)$$

The thin pencil beams of neutrinos for experiments will be produced from long straight sections in either the collider ring or a ring dedicated to neutrino physics. From relativistic kinematics, the forward hemisphere in the muon rest frame will be boosted, in the lab frame, into a narrow cone with a characteristic opening half-angle, $\theta_\nu$, given in obvious notation by

$$\theta_\nu \simeq \sin\theta_\nu = 1/\gamma = \frac{m_\mu}{E_\mu} \simeq \frac{10^{-4}}{E_\mu(\text{TeV})}. \quad (2)$$

For the example of 250 GeV muons, the neutrino beam will have an opening half-angle of approximately 0.4 mrad. The final focus regions around collider experiments are important exceptions to equation 2 since the muon beam itself will have an angular divergence in these regions that is large enough to significantly spread out the neutrino beam.

For TeV-scale neutrinos, the neutrino cross-section is approximately proportional to the neutrino energy, $E_\nu$. The charged current (CC) and neutral current (NC) interaction cross sections for neutrinos and antineutrinos have numerical values of [1]:

$$\sigma_{\nu N} \text{ for } \begin{pmatrix} \nu\_CC \\ \nu\_NC \\ \overline{\nu}-CC \\ \overline{\nu}-NC \end{pmatrix} \simeq \begin{pmatrix} 0.72 \\ 0.23 \\ 0.38 \\ 0.13 \end{pmatrix} \times \frac{E_\nu}{1\text{ TeV}} \times 10^{-35} \text{ cm}^2. \quad (3)$$

These cross sections are easily converted into approximate experimental event rates for the example of a 250+250 GeV collider with a 200 meter straight section and the example design parameters used for this workshop. For a general purpose detector subtending the boosted forward hemisphere of the neutrino beam:

$$\text{Number of } \begin{pmatrix} \nu_\mu - CC \\ \nu_\mu - NC \\ \overline{\nu}_e - CC \\ \overline{\nu}_e - NC \end{pmatrix} \text{ events/yr} \simeq \begin{pmatrix} 2.6 \\ 0.8 \\ 1.4 \\ 0.5 \end{pmatrix} \times 10^7 \times l[\text{g.cm}^{-2}], \quad (4)$$

and for a long baseline detector in the center of the neutrino beam:

$$\text{Number of } \begin{pmatrix} \nu_\mu - CC \\ \nu_\mu - NC \\ \overline{\nu}_e - CC \\ \overline{\nu}_e - NC \end{pmatrix} \text{ events/yr} \simeq \begin{pmatrix} 1.4 \\ 0.4 \\ 0.7 \\ 0.2 \end{pmatrix} \times 10^7 \times \frac{\text{M[kg]}}{(\text{L[km]})^2}. \quad (5)$$

These event rates are several orders of magnitude higher than in today's neutrino beams from accelerators.

# POTENTIAL RADIATION HAZARD

The neutrino fluxes strong enough to constitute a potential off-site neutrino radiation hazard [2]. The problem comes where the neutrinos from the collider ring exit the Earth's surface and their interactions initiate showers of ionizing particles in people and their surroundings. The neutrino interaction cross section is tiny but this is greatly compensated by the huge numbers of neutrinos. A simple but conservative order-of-magnitude calculation [2] predicts the following maximum radiation dose downstream from a straight section of length $S$:

$$\frac{\text{Radiation dose}}{\text{U.S. Fed. limit}} \simeq 0.3 \times \left(\frac{S}{\text{collider depth}}\right) \times \left(\frac{\text{muon current}}{6 \times 10^{20} \ \mu^-/\text{yr}}\right) \times \left(\frac{E_\mu}{250 \ \text{GeV}}\right)^3 \quad (6)$$

The first bracket on the right hand side is valid when the collider ring is not tilted and the land around the collider is flat enough to assume the approximation of a spherical Earth. In practice, a dedicated ring for neutrino physics would almost certainly be tilted downwards towards a long-baseline neutrino experiment of order 1000 km away. Obviously, the radiation hazard would be much reduced for this orientation. The muon current of $6 \times 10^{20}$ is the example design specification for this workshop.

The cubic dependence on energy means that neutrino radiation is a serious design constraint for colliders at the TeV center-of-mass (CoM) energy scale and above. In practice, this is partially compensated by the increasing luminosity per unit current that can be obtained at higher energies, and high luminosity designs with decreased muon current appear to be feasible to at least the 10 TeV CoM energy scale [3].

# A GENERAL PURPOSE NEUTRINO DETECTOR

Figure 1 is an example of the sort of high rate general purpose neutrino detector that would be well matched to the intense neutrino beams. Note the contrast with the kilotonne-scale calorimetric targets used in today's high rate neutrino experiments.

The neutrino target is a stack of CCD tracking planes. A target vessel containing hydrogen could equally well be used, alternating between runs with protium and deuterium. The target is 1 meter long and has a 10 cm radius, which matches the beam radius at approximately 200 meters from production for a 250 GeV muon beam. Besides providing the mass for neutrino interactions, such a detector allows precise reconstruction of the event topologies from charged tracks, including event-by-event vertex tagging of those events containing charm or beauty hadrons (or tau leptons). The vertexing geometry can be made essentially optimal for vertex tagging, with 3.5 micron hit resolution or better (this is the resolution of the CCD's at the SLD detector), shorter extrapolations to secondary vertices than is possible in collider experiments and negligible acceptance inefficiency. It is reasonable to expect almost 100 percent efficiency for b tagging, perhaps 70 to 90 percent efficiency for charm tagging (depending on the tag purity and how 1-prong decays are treated), and excellent discrimination between b and c decays.

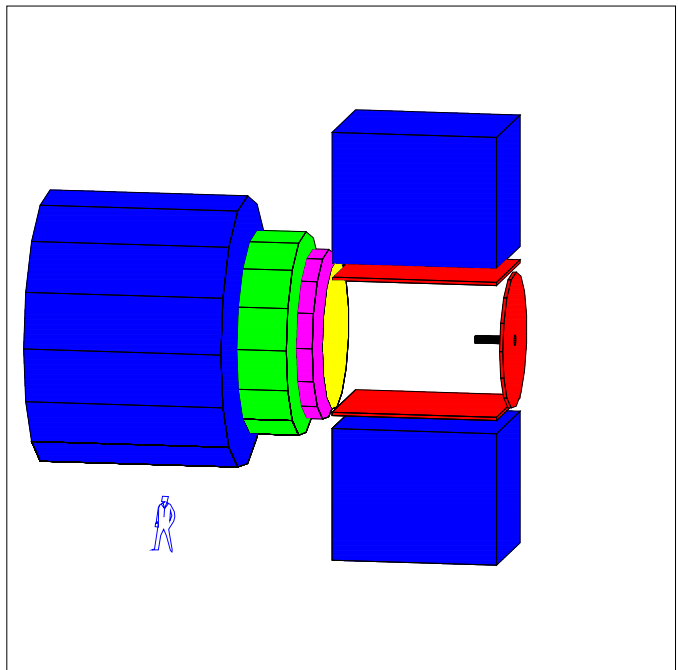

**FIGURE 1.** Example of a general purpose neutrino detector. A human figure in the lower left corner illustrates its size. The neutrino target is the small horizontal cylinder at mid-height on the right hand side of the detector. Its radial extent corresponds roughly to the radial spread of the neutrino pencil beam, which is incident from the right hand side. Further details are given in the text.

The optimization of the number of planes and their thickness is a trade-off between increased detector mass and degradation of the tracker performance due to multiple coulomb scattering and electromagnetic and hadronic secondary interactions in the tracker. For example, 750 planes of 300 micron thick silicon CCD's corresponds to a mass per unit area of approximately 50 g.cm$^{-2}$, about 2.5 radiation lengths and 0.5 interaction lengths. This seems reasonable for a general purpose detector, while detector masses perhaps up to an order of magnitude larger might be considered for a very high rate experiment dedicated to rare processes. In the neutrino beam assumed for equation 4, such a general purpose detector would record a very healthy $2 \times 10^9$ CC interactions per year and even higher statistics are obviously possible using a higher mass target, higher energy neutrino beam and/or a dedicated muon ring.

In figure 1 the target is surrounded by a gas-based tracking subdetector, using a time projection chamber (TPC) geometry with a vertical magnet field provided by a dipole magnet with poles above and below the tracker. The characteristic dE/dx signatures from the tracks would identify each charged particle. Further particle

ID is provided by the Cherenkov photons produced in the gas tracker, which are reflected by a spherical mirror at the downstream end of the tracker and focused onto a read-out plane at the upstream end of the target. If desired, a third level of electron ID could be provided by transition radiation detectors (TRD's) behind the mirror. The mirror is backed by electromagnetic and hadronic calorimeters and, lastly, by iron-core toroidal magnets for muon ID.

# NEUTRINO INTERACTIONS AND THEIR EXPERIMENTAL INTERPRETATION

The dominant interaction of TeV-scale neutrinos is deep inelastic scattering (DIS) off nucleons (i.e. protons and neutrons). Scattering off atomic electrons also occurs, but at a rate which is lower by three orders of magnitude. There are 2 types of DIS: neutral current (NC) and charged current (CC) scattering, as follows.

In neutral current (NC) scattering, the neutrino is deflected by a nucleon (N) and loses energy with the production of several hadrons (X):

$$\nu + N \rightarrow \nu + X, \tag{7}$$

This comprises about 25 percent of the total cross section and is interpreted as elastic scattering off one of the many quarks inside the nucleon through the exchange of a virtual neutral Z boson:

$$\nu + q \rightarrow \nu + q. \tag{8}$$

Charged current (CC) scattering is similar to NC scattering except that the neutrino turns into its corresponding charged lepton:

$$\nu + N \rightarrow l^- + X,$$
$$\overline{\nu} + N \rightarrow l^+ + X, \tag{9}$$

where $l$ is an electron/muon for electron/muon neutrinos. At the more fundamental quark level a charged W boson is exchanged with a quark ($q$), which is turned into another quark species ($q'$) whose charge differs by one unit.

$$\nu + q \rightarrow l^- + q'. \tag{10}$$
$$\overline{\nu} + q' \rightarrow l^+ + q. \tag{11}$$

The relativistically invariant quantities that are routinely extracted in DIS experiments are 1) Feynman $x$, the fraction of the nucleon momentum carried by the struck quark, 2) the inelasticity, $y = E_{\text{had}}/E_\nu$, which is related to the scattering angle of the neutrino in the neutrino-quark CoM frame, and 3) the momentum-transfer-squared, $Q^2 = 2M_{proton}E_\nu xy$.

Equation 9 shows that the interactions of muon- and electron-flavor neutrinos are essentially identical for NC interactions but easily distinguishable in CC interactions by the flavor of the final state charged lepton. Similarly, the charge of the final state lepton clearly distinguishes the CC interactions of neutrinos from antineutrinos. Even the NC interactions of neutrinos and antineutrinos are distinguishable on a statistical basis because their $y$ distributions differ for reasons involving the differing helicity states of neutrinos and antineutrinos.

The final state quark always "hadronizes" at the nuclear distance scale, combining with quark-antiquark pairs to produce the several hadrons seen in the detector. It is an important aspect of the MURINE that the hadronic final state will retain some memory of the final state quark flavor since this quark must still exist in one of them. This is a particularly good signature for the neutrino-induced production of the heavy charm and beauty quark flavors since these flavors of quark-antiquark pairs are very rarely produced during hadronization. Hence, the observation of the characteristic displaced vertex of a charm or beauty hadron in an event gives a fairly reliable identification of c or b production from the neutrino interaction.

Some statistically based flavor tagging will be available even when the final state quark is one of the lighter u, d or s quarks. This comes from the so-called "leading particle effect" and is currently used, for example, in analyses at the LEP experiments. The hadron containing the struck quark is known to be more energetic on average than the others, so the quark content of the most energetic, i.e. "leading", hadron provides a potential tag for the final state quark flavor.

A particularly good example is those events where most of the hadronic energy is concentrated in a kaon–anti-kaon pair from the $\phi(1020)$ resonance. The $\phi(1020)$ consists of an s quark and s antiquark, so such events can reliably be interpreted as a final state strange quark (or antiquark) that has subsequently combined with its strange antiparticle during hadronization.

The relatively high purity $\phi(1020)$ tag is one of several possible final state tags. Events with most of the hadronic energy concentrated in a single kaon provide another tag for a strange quark final state which has lower purity but higher efficiency and, clearly, the complementary sample of events with no high energy kaons will be enriched in interactions with u and d final state quarks.

As an important technical detail, it appears that the efficiency and purity of the various u, d and s tags should be measurable with little uncertainty by taking advantage of the various possible calibration event samples. For example, neutrino interactions with the u and d valence quark components of protons and neutrons (which dominate at high x) will produce accurately predictable fractional compositions of u, d and/or s quark final states. Also, as a more general handle, charge conservation dictates that the hadron containing the final state quark from neutrino (anti-neutrino) CC interactions must be either neutral or positively (negatively) charged, so the production of wrong-sign high energy hadrons gives a direct measure of the impurity of each charged hadron tag. To the extent that such calibration analyses can remove systematic uncertainties in the final state tagging of light quarks, the only price to pay for the statistical nature of these tags will be an effective dilution of the event statistics.

# PHYSICS OPPORTUNITIES

Neutrino interactions are interesting both in their own right and as probes of the quark content of nucleons, so a MURINE has wide-ranging potential to make advances in many areas of research in elementary particle physics. There is insufficient space to do justice to all the physics possibilities and it actually seems almost easier to list the areas that can't be studied at a MURINE! Significant exceptions are some types of rare decay processes, studies involving the decay of b hadrons and the direct production (as opposed to virtual production) of particles heavier than b hadrons.

The first part of this section gives a discussion of one area the author finds particularly interesting – measurements involving the CKM quark mixing matrix. Briefer overviews are then given of several other areas of physics. These include tests of nucleon structure and QCD, electroweak measurements, neutrino oscillations, the search for exotic particles and, finally, studies of charmed hadrons.

For a benchmark event sample, CC statistics of $10^{10}$ events are assumed for the general purpose detector, corresponding to 5 years running with a target mass of 50 g.cm$^{-2}$. This statistics is several ordersof magnitude higher than today's data samples. In all cases, any quantitative predictions for the expected precision of analyses are little more than guesses based on cursory examinations of quark distributions etc. and rough comparisons with analyses of existing neutrino experiments. Clearly, further work in this area is both needed and interesting.

## CKM Quark Mixing Matrix

There is considerable theoretical interest in the mixture of final state quarks produced in CC interactions. The struck quark can be converted into any of the three final state quarks that differ by one unit of charge: a down (d), strange (s), or bottom(b) quark can be converted into an up (u), charmed (c), or top (t) quark and vice versa. In practice, production of the heavy top quark is kinematically forbidden at these energies and the production of other quark flavors is influenced by their mass. Beyond this, the Standard Model (SM) predicts the probability for the interaction to be proportional to the absolute square of the appropriate element in the so-called Cabbibo-Kobayashi-Maskawa (CKM) quark mixing matrix. The squared terms are given in table 1 [4].

The 4 independent parameters in the underlying 3-by-3 unitary CKM matrix are as fundamental as the masses of elementary particles and, like particle masses, their values are phenomenological parameters to be determined by experiment rather than being predicted by the SM. One of the parameters is a complex phase that is postulated as an explanation for CP violation: the intriguing experimental phenomenon that particles may have tiny deviations from the properties that mirror those of their antiparticles. This phase is poorly constrained by experiment and, in fact, experimental measurements involving the CKM matrix are sufficiently difficult that even the unitary form of the matrix as predicted by the SM is not particularly well established. Improved measurements involving the CKM matrix will test the SM hypotheses and, speculatively, the values and pattern of the parameters might even help provide insights towards some deeper future theory that explains such mysteries of the SM as why nature chooses to have 3 quark generations.

**TABLE 1.** Quark mixing probabilities. Threshold suppression due to quark masses has been neglected. In practice, this will reduce the mixing probabilities to the heavier c and b quarks to below the values given in the table and will prevent any mixing to the top quark.

|   | d | s | b |
|---|---|---|---|
| **u** | 0.95 | 0.05 | $1 \times 10^{-5}$ |
| **c** | 0.05 | 0.95 | 0.002 |
| **t** | $1 \times 10^{-4}$ | 0.001 | 1 |

**TABLE 2.** Percentage uncertainties in quark mixing probabilities. The two terms in brackets have not been measured directly from tree level processes.

|   | d | s | b |
|---|---|---|---|
| **u** | 0.1% | 1.6% | 50% |
| **c** | 15% | 35% | 15% |
| **t** | (25%) | (40%) | 30% |

The current percentage uncertainties in the 9 mixing probabilities are given in table 2 [4]. One of the CKM mixing probabilities – that between the d and c quarks – is already best measured in neutrino-nucleon scattering. It appears that the new realm of neutrino experiments at muon colliders could make a dominant contribution to this field, with measurements of perhaps 4 of the 9 mixing probabilities that should be far better than possible in any other type of experiment. The relevant transitions are that between u and b quarks, and those between c quarks and d, s or b quarks.

The d-to-c transition is the cleanest of the 4 measurements, with statistics of order $10^8$ events, a well measured distribution of valence d quarks at high x and clean event-by-event vertex tagging of the charm final state. The threshold effect due to the charm quark mass must also be modeled, but this should be little problem with such statistics and the measurement precision should reach the parts-per-mil level.

Measurement of the s-to-c transition will also benefit from of order $10^8$ statistics and clean charm tagging. The major difficulty here will be an incomplete knowledge of the initial strange sea distribution, and presumably this will estimated using neutral current interactions with a resonant $\phi(1020)$ final state, as explained in the preceding section. The resulting measurement accuracy might be at the percent level. As in analyses of today's neutrino experiments, the d and s contributions to the charm event sample will in practice be separated from one another in a fit involving the x distributions etc. and with the charm quark mass as a fitted parameter.

A similar analysis will be applied for the two transitions with a b quark in the final state: u-to-b and c-to-b. The statistics for these two processes will be perhaps of order $10^4$ events, depending on the energy of the muon storage ring and the consequent threshold suppression due to the b quark mass. (The b quark mass is approximately 5 GeV compared to about 1.5 GeV for the charm quark. These masses would be measured with unique precision in a MURINE.) Besides the threshold suppression, other reasons for the reduced statistics are that the u-to-b transition has a low mixing probability – about $10^{-5}$ – while the initial charm quark distribution (either an intrinsic charm sea or charm production through higher order Feynman diagrams) will much smaller than those of the lighter quark.

Given the nearly optimal vertexing geometry possible at a MURINE, the vertexing discrimination between b and c hadrons should be able to separate out much of the fractionally small b hadron event sample. (Technically, this can be done by requiring a secondary vertex with a total invariant mass larger than possible for a charm hadron and/or using other "topological vertexing" signatures.) If this is so then the u-to-b analysis should be a relatively straightforward copy of the d-to-c analysis, and the precision of the measurement may approach the statistical limit of around 1 percent. In contrast, the c-to-b transition has the additional challenge of estimating the initial charm quark spectrum using NC charm events and/or CC scattering involving the c-to-s transition, and this may well limit the measurement accuracy to the few percent level.

In summary, a first look indicates the opportunity for tremendous improvements in measuring the quark mixing matrix. More detailed studies are clearly desirable.

## Structure Functions and QCD

Another major motivation for MURINE's is the potential for greatly improved measurements of nucleon structure functions (SF) and the consequent tests of quantum chromodynamics (QCD) – the theory of the strong interaction that is widely accepted for its elegance and simplicity but which has not been experimentally verified at the level of the electroweak theory.

Stated simply, quark SF's are the momentum distributions of the quarks inside the nucleon. They are assumed to be universal properties of nucleons which can also be measured and compared at charged lepton scattering experiments, either at a fixed target or a collider (of which HERA is the only operating example). QCD predicts a weak dependence of the SF on the energy transferred in the scattering interaction and the improved verification of this prediction and of various "quark sum rules" at a MURINE will constitute some of the best tests for QCD and also provide one of the best measurements of the coupling strength parameter of the strong interaction, $\alpha_s$.

Neutrino scattering experiments differ from their charged lepton counterparts in that the scattering proceeds predominantly through the exchange of W and Z bosons, rather than photons. This gives neutrino scattering experiments two important advantages:

1. the heavy mass scale of the W and Z bosons naturally produces hard scattering interactions that can be well treated by assuming scattering off a quasi-free quark with QCD corrections. In contrast, charged lepton scattering suffers from a huge background of soft photon interactions that provide experimental rate problems and are not very amenable to analysis using perturbative QCD.

2. the relatively evenly divided event statistics from $W^+$, $W^-$ and Z interactions provides much more discrimination between the quark flavors than is possible with a single scattering probe. For example, neutrino experiments have the unique capability to disentangle the valence quark SF without even requiring identification of the final state quarks.

Beyond the intrinsic advantages of neutrino interactions, new capabilities will arise from a MURINE with its precision high rate detector. For the first time, the precisely known beam spectra and detailed reconstruction of the hadronic final state will allow precise SF from NC interactions, while measurement of CC event kinematics will be overconstrained and hence lead to extremely precise CC structure functions with minimal systematic uncertainties. Further, the new capability to identify the final state quark will enable flavor-by-flavor SF measurements. The use of several different target materials will also allow studies of nuclear effects, polarized structure functions and separate SF for protons and neutrons. Finally, the reconstruction of the hadronic final state should, for the first time in lepton scattering experiments, allow high statistics studies of gluon splitting and jet topologies similar to those done, e.g., on the LEP sample of Z decays.

In summary, a MURINE should greatly enrich and extend our knowledge of nucleon structure functions and might well be the best single experiment of any sort for the examination of perturbative QCD.

# Electroweak Parameters

Neutrino physics has had an important historical role in measuring the electroweak mixing angle, which is simply related to the mass ratio of the W and Z intermediate vector bosons:

$$\sin^2 \theta_W \equiv 1 - \left(\frac{M_W}{M_Z}\right)^2. \tag{12}$$

(To be precise, this equation is the Sirlin on-shell definition of $\sin^2 \theta_W$ that is conventionally used in neutrino physics. Several other definitions of $\sin^2 \theta_W$ differ from equation 12 by small amounts are used for convenience in other types of experimental measurement.

Now that $M_W$ has been precisely measured at LEP, measurements of $\sin^2 \theta_W$ in neutrino physics can be directly converted to predictions for the W mass. The comparison of this prediction with direct $M_W$ measurements in collider experiments constitutes a precise prediction of the SM and a sensitive test for exotic physics modifications to the SM. Today's neutrino measurements correspond to an uncertainty in $M_W$ of $\Delta M_W = 180$ MeV [5], with contributions from both event statistics and experimental and theoretical systematic uncertainties.

A crude estimate of the predicted uncertainty in $M_W$ from a MURINE analysis was made by reviewing each of the uncertainties in a contemporary $\sin^2 \theta_W$ analysis from the CCFR collaboration [6]. This exercise was made more difficult by the enormously improved experimental conditions and the consequently large extrapolations in experimental accuracy.

The huge statistics at a MURINE should reduce the $M_W$ statistical uncertainty to only a couple of MeV and, given the high quality beam and high performance detector, the experimental uncertainties may well be reducable to a similar level. Further, all of the large sources of large theoretical systematic uncertainties in the CCFR analysis – such as the threshold suppression of charm production, longitudinal SF and the heavy flavor content of the nucleon sea – should be controllable through direct measurements in a MURINE. Radiative corrections, which can't be measured directly, were assigned an uncertainty corresponding to only about 5 MeV. Based on these observations, it appears that MURINE's have the potential for precisions more than an order of magnitude better than today's $\sin^2 \theta_W$ measurements. Thus, W mass predictions with precisions of order 10 MeV seem achievable. This is comparable with the projected best direct measurements from future collider experiments.

Another electroweak measurement of interest for MURINE's is the $\sin^2 \theta_W$ measurement from neutrino scattering off electrons in the target:

$$\begin{aligned} \nu\, e^- &\to \nu\, e^- \quad \text{(NC)} \\ \nu\, e^- &\to l\, \nu_e \quad \text{(CC)}. \end{aligned} \tag{13}$$

This is a theoretically clean process that turns out to give interesting physics information orthogonal to the measurement from neutrino-quark scattering. However, it has been experimentally less accessible because the cross section is lower by three orders of magnitude. Current measurements are limited both by statistics and from the reconstruction limitations of today's high mass neutrino targets, both of which would be enormously improved at a MURINE.

Other new and/or greatly improved electroweak measurements could be performed peripherally to these major analysis topics. These include determinations of the left-handed and right-handed neutrino-quark couplings for each quark flavor, and a precise check of the SM prediction for the inverse muon decay cross section.

## A  Neutrino Oscillations

A neutrino property that is currently drawing much interest is the question of whether neutrinos have a non-zero mass. If they do then it is possible that the 3 neutrino flavors mix to produce neutrino oscillations that can be observed using a neutrino beam. The probability for an oscillation between two of the flavors is given by [7]:

$$\text{Oscillation Probability} = \sin^2\theta \times \sin^2\left(1.27\frac{\Delta m^2[\text{eV}^2].L[km]}{E_\nu[GeV]}\right). \quad (14)$$

The first term gives the mixing strength and the second term gives the distance dependence. The discovery potential for a neutrino experiment is conventionally expressed as a characteristically shaped [7] region in an exclusion/discovery plot of $\Delta m^2$ vs. $\sin^2\theta$. The projections of this plot on the x and y axes give the limits on mixing strength for most favorable mass difference and on the mass difference for full mixing, respectively. These two limits are now crudely estimated, with both estimates applying generically to all 3 possible mixings between 2 flavors:

$$\nu_e \leftrightarrow \nu_\mu,$$
$$\nu_e \leftrightarrow \nu_\tau,$$
$$\nu_\mu \leftrightarrow \nu_\tau. \quad (15)$$

The best mass difference limit would come from a long baseline experiment. The background-free 90 % confidence limit of 2.3 oscillated events becomes a reasonable approximation with a low enough event rate and with cuts applied so that only reliably tagged oscillations are accepted. If it is crudely assumed that this corresponds to a 10 % tagging efficiency for oscillated events then 5 years running at a 10 kilotonne long-baseline experiment is easily found to give the following order-of-magnitude mass limit for full mixing:

$$\Delta m^2|_{min} \sim O(10^{-4})\ eV^2, \quad (16)$$

independent of the distance to the detector. This is more than an order of magnitude better than any proposed accelerator or reactor experiments for $\nu_\mu \leftrightarrow \nu_\tau$ and $\nu_e \leftrightarrow \nu_\tau$, and competitive with the best such proposed experiments for $\nu_e \leftrightarrow \nu_\mu$.

The limit on mixing strength would benefit not only from the neutrino beam properties but also from the high performance of the general purpose neutrino detector. The overconstrained event kinematics for CC events (see the subsection on SF's) should greatly benefit the selection of a clean potential oscillation sample. The search for oscillations to tau neutrinos would also take advantage of the exceptional vertex tagging abilities of such a detector. The huge event samples would allow an analysis strategy of cutting hard on event kinematics to keep only very well reconstructed events with an unambiguous primary lepton. Assuming that 1 in 100 oscillating events would pass the cuts and requiring a signal of 10 such events corresponds to a mixing probability sensitivity for $10^{10}$ events of:

$$\sin^2\theta|_{min} \sim O(10^{-7}). \tag{17}$$

This would be a unique sensitivity for each of the three possible oscillations – orders of magnitude better than in any other current or proposed experiment.

## Searches for Exotic Particles

It is clear that MURINE's will offer expanded opportunities for searches for new types of exotic particles that couple to neutrinos, such as some hypothesized types of neutral heavy leptons.

## B  Charm Physics

Interestingly, MURINE's should be rather impressive factories for the study of charm, with a clean, well reconstructed sample of several times $10^8$ charmed hadrons produced in $10^{10}$ neutrino interactions. The particle ID and energy/momentum capabilities of the detector should facilitate the full reconstruction of a good fraction of the decay final states, and the precision vertexing should give accurate event-by-event lifetime information. It is of particular importance to oscillation and CP studies that the production sign of the charm quark is tagged by the final state lepton charge in CC interactions:

$$\begin{aligned}\nu q &\to l^- c \\ \overline{\nu} q' &\to l^+ \overline{c}.\end{aligned} \tag{18}$$

There are several interesting physics motivations for charm studies at a MURINE [8]. Measurement of charm decay branching ratios and lifetimes are useful for both QCD studies and for the theoretical calibration of the physics analyses on B hadrons. Charm decays also provide a "clean laboratory" to search for exotic physics contributions, with the SM predicting 1) tiny branching ratios for rare decays, 2) small CP asymmetries and 3) slow $D^0 \to \overline{D^0}$ oscillations, with only of order 1 in $10^4$ oscillating before decay. In fact, $D^0 - \overline{D^0}$ mixing [9] has yet to be observed, and it is quite plausible that a MURINE would provide the first observation.

# SUMMARY

The intense neutrino beams at muon collider complexes should usher in an exciting new era of neutrino physics experiments. It has been shown that great advances are to be expected in traditional areas of neutrino physics and elsewhere.